# Sounds Synthesis with Slime Mould of *Physarum Polycephalum*

Eduardo R. Miranda[1], Andrew Adamatzky[2] and Jeff Jones[2]

[1] Interdisciplinary Centre for Computer Music Research (ICCMR), University of Plymouth, Plymouth, PL4 8AA UK; eduardo.miranda@plymouth.ac.uk

[2] Unconventional Computing Centre, University of the West of England, Bristol, BS16 1QY UK; andrew.adamatzky@uwe.ac.uk

**Abstract**

*Physarum polycephalum* is a huge single cell with thousands of nuclei, which behaves like a giant amoeba. During its foraging behaviour this plasmodium produces electrical activity corresponding to different physiological states. We developed a method to render sounds from such electrical activity and thus represent spatio-temporal behaviour of slime mould in a form apprehended by humans. We show to control behaviour of slime mould to shape it towards reproduction of required range of sounds.

## 1 Introduction

Our research is concerned with the application of novel computational paradigms implemented on biological substrates in the field of computer music. Computer music has evolved in tandem with the field of Computer Science. Computers have been programmed to produce sounds as early as the beginning of the 1950's. Nowadays, the computer is ubiquitous in many aspects of music, ranging from software for musical composition and production, to systems for distribution of music on the Internet. Therefore, it is likely that future developments in fields such as Bionic Engineering will have an impact in computer music applications.

Research into novel computing paradigms in looking for new algorithms and computing architectures inspired by, or physically implemented on, chemical, biological and physical substrates (Calude et al. 1998, Adamatzky and Teuscher 2005, Adamatzky *et al.* 2007). We are interested in exploring ways in which such unconventional modes of computation may provide new directions for future developments in Computer Music.

Prototypes of novel computational devices that have been recently developed include DNA computers, reaction-diffusion chemical computers, molecular machines and bacterial computers. However, these are costly to build and maintain. They require sophisticated laboratory resources and highly specialist training of personnel to conduct experiments; see discussion in (Adamatzky 2010). Conversely, the plasmodial slime mold *Physarum polycephalum* is a biological computing substrate, which requires comparatively less resources and is more cost-effective than those prototypes mentioned above.

*Physarum polycephalum* is a single cell with multiple nuclei. When placed on a substrate with scattered sources of nutrients it forms a network of protoplasmic tubes connecting the sources (Figure 1). The plasmodium optimizes its protoplasmic network for efficient utilisation of resources and flow of intra-cellular components (Nakagaki *et al*. 2000). Its optimal foraging and reaction to attracting (e.g., food and humidity) and repelling (e.g., light and salt) sources makes it an ideal candidate for research into biological unconventional computers. It has been already demonstrated that computing devices based on this plasmodium were capable of solving difficult classic computational problems such as: approximation of shortest path (Nakagaki *et al*. 2001), planar proximity graphs (Adamatzky 2009a), Voronoi diagram (Shirakawa *et al*. 2010), execution of basic logical operations (Tsuda et al. 2004, Adamatzky 2009b), spatial logic and process algebra (Schumann and Adamatzky 2009). *Plasmodium of Physarum* is experimentally proved to be an original and efficient micromanipulator controlled by light (Adamatzky and Jones 2008). For an overview of computing devices built with *Physarum polycephalum* please refer to (Adamatzky 2010).

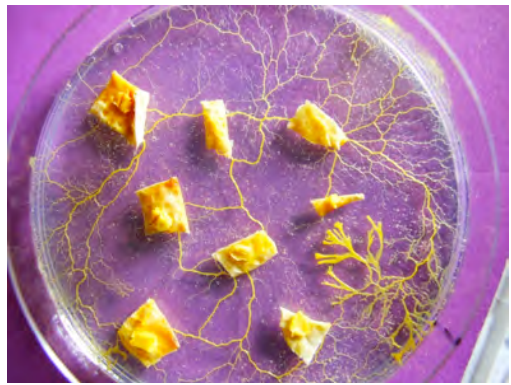

**Figure 1**. Plasmodium of *Physarum polycephalum* in a Petri dish.

As all living creatures, *Physarum polycephalum* produces electrical activity. The movement of intra-cellular components inside the plasmodium's body and its protoplasmic tubes, and migration of the plasmodium over a substrate, produce electricity that can be measured with electrodes (Kamiya and Abe 1950, Kishimoto 1958). Recently Adamatzky and Jones (2010) studied the electrical activity of such plasmodium and they found patterns of electrical activity, which uniquely characterise the plasmodium's spatial dynamics and physiological state. Different measurements of electrical potentials, or voltages, indicated when the plasmodium occupied and when it left specific sites of its substrate. They also indicated when the organism functioned properly, when it was in a state of distress, and when it entered in hibernation mode.

In this paper we report a method we developed to render sounds from such electrical activity. At this stage of this research we are not concerned with studying the computational properties of the plasmodium. Rather, we are interested in understanding its behaviour and in the application of this behaviour to build *bionic musical instruments*. *Physarum polycephalum* is interesting because it behaviour can be controlled to produce variations of its electrical activity (e.g., by placing attractors in the space) and consequently variations on the sounds that it produces.

By way of related research into bionic musical instruments we cite the development of a granular synthesiser using models of reaction-diffusion chemical computing (Miranda 1995,

Serquera and Miranda 2010) and a method to sonify of the behaviour of *in vitro* neural networks (Miranda *et al*. 2009). As far as we are aware, we are pioneers in applying the *Physarum polycephalum* slime mould in computer music.

The rest of the paper is structured as follows. Firstly we introduce our method to culture the plasmodium and record its electrical activity. Then, we describe a typical experiment and introduce our method to synthesise sounds from the plasmodium electrical activity. Two examples of sounds are discussed, followed by concluding remarks.

## 2 Experimenting with Physarum

The electrical activity of *Physarum polycephalum* is recorded with an ADC-16 High Resolution Data Logger, by Pico Technology, UK. In each experiment the plasmodium is cultured in a Petri dish of 9 cm of diameter. In each Petri dish, we place one reference electrode and a number of measurement electrodes and cover them with blobs of non-nutrient agar gel; the naked part of a coated wire acts as an electrode here. At the beginning of each run, a piece of plasmodium is placed on the reference electrode's agar blob (Figure 2). The plasmodium feeds on oat flakes. Therefore, an oat flake is placed on top of each agar blob to act as nutrients, which will eventually attract the plasmodium to colonise the measurement electrodes. Agar blobs do not touch each other. They are separated by a strip of non-conductive plastic placed at the bottom of a Petri dish.

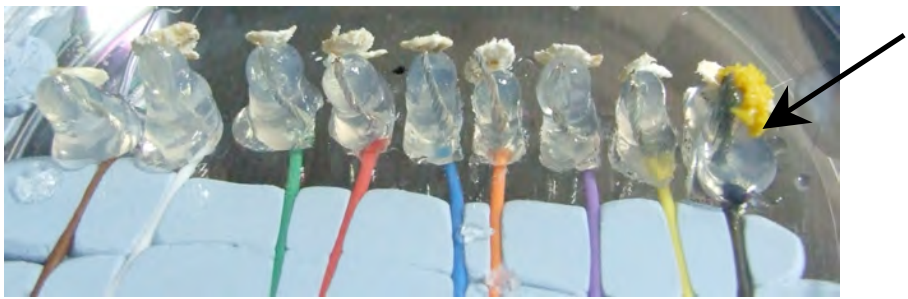

**Figure 2**: Photo of an example setup. The reference electrode is on the right side. The other 8 electrodes are measurement electrodes. They are covered with blobs of non-nutrient agar and oat flakes placed on top of each. At the beginning, a piece of plasmodium is placed on top of agar blob covering the reference electrode only (indicated by the dashed circle).

The plasmodium spreads to the measuring electrodes slowly: in average, each run takes 1 week to complete. We record the voltages from the measurement electrodes every second. In practice we sample 100 measurements in one second and then these values are averaged. Furthermore, in order to compress data worth of several days of activity into data suitable to produce a few minutes of sound, we process the electric potentials from electrodes $e_1 \ldots e_8$ as follows: measurements $e_1^t \ldots e_8^t$ taken at time step *t* are taken into account only if at least *N* electrodes presented a change in their electric potential. Otherwise this measurement is skipped. That is, $\sum_{i=1}^{8} \chi(|e_i^t - e_i^{t-1}| > 0.4) > 4$ where $\chi(C) = 1$ if predicate $C$ is true and 0 otherwise. Also, voltage values were capped in a range between -40mV to 40mV, which corresponds to a normal range of plasmodium activity, and yet this removes potential interference from nearby electrical equipment. This value is subsequently scaled by $S$ in order match the requirements of the synthesis algorithm; for instance, $N = 5$ and $S = 20$ were used to produce the sound examples of this paper.

## 3 Behaviour and Control

Considering the setup shown in Figure 2, Figures 3 and 4 illustrate a typical example of plasmodium activity. The plasmodium gradually proliferates from its initial position on the reference electrode onto the other electrodes (Figure 3). Typically this takes place in the course of approximately three days. The colonisation of an electrode produces a characteristic pattern of voltage dynamics. At first the electrode being colonised registers a rise in its voltage by up to 20 mV (Figure 4). Then, a drop follows this rise, which sometimes can be as large as -40 mV (Figure 4). In Figure 4 the voltages are plotted obeying the order in which the colonisation took place from the right (electrode 1) to the left side of Figure 3 (electrode 8); note that electrode 6 has not been colonised in this example. This might have been caused by a number of reasons; e.g., the site might have been infected by a bacteria or fungi that did not incite the appetite of the plasmodium.

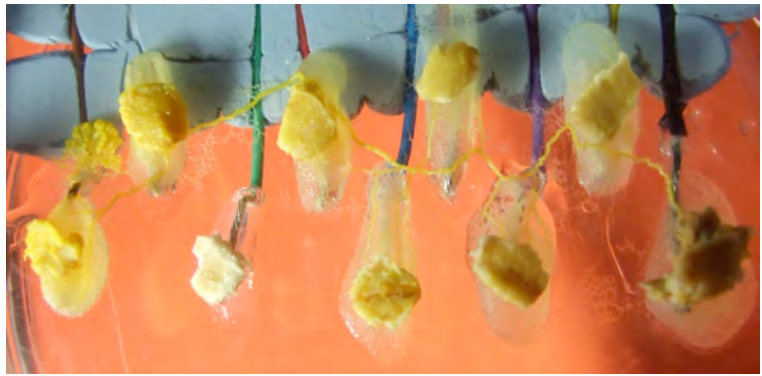

**Figure 3:** The plamodium proliferates from its initial position on the reference electrode (first on the right side of the figure), towards the other 8 measurement electrodes. Note that the 6$^{th}$ measurement electrode has not been colonised by the plasmodium.

Eventually, the plasmodium abandons the agar blobs as they start to dry and/or nutrients are drained. A gradual decrease in voltage is registered when the plasmodium abandons an agar blob. When blobs dry and/or nutrients finish, the plasmodium gets into a state of hibernation, forming what is referred to as *sclerotium*. The sclerotium is characterised by a positive electric potential. This can be seen in channel 2, starting at approximately 25,000 seconds, Figure 4.

The plasmodium's voltage between stages of colonisation and hibernation is usually highly dynamic and complex. It represents the interaction of many travelling waves of excitation and contractile waves, including the formation and annihilation of bio-chemical oscillators, branching of plasmodium tree, and elimination of some protoplasmic tubes. Propagation of strong contractile waves can be seen as a series of electric impulses detected in the electrodes chain. This can be seen in channels 3, 4 and 5 in Figure 4; note the anti-phase oscillations of voltages in these channels.

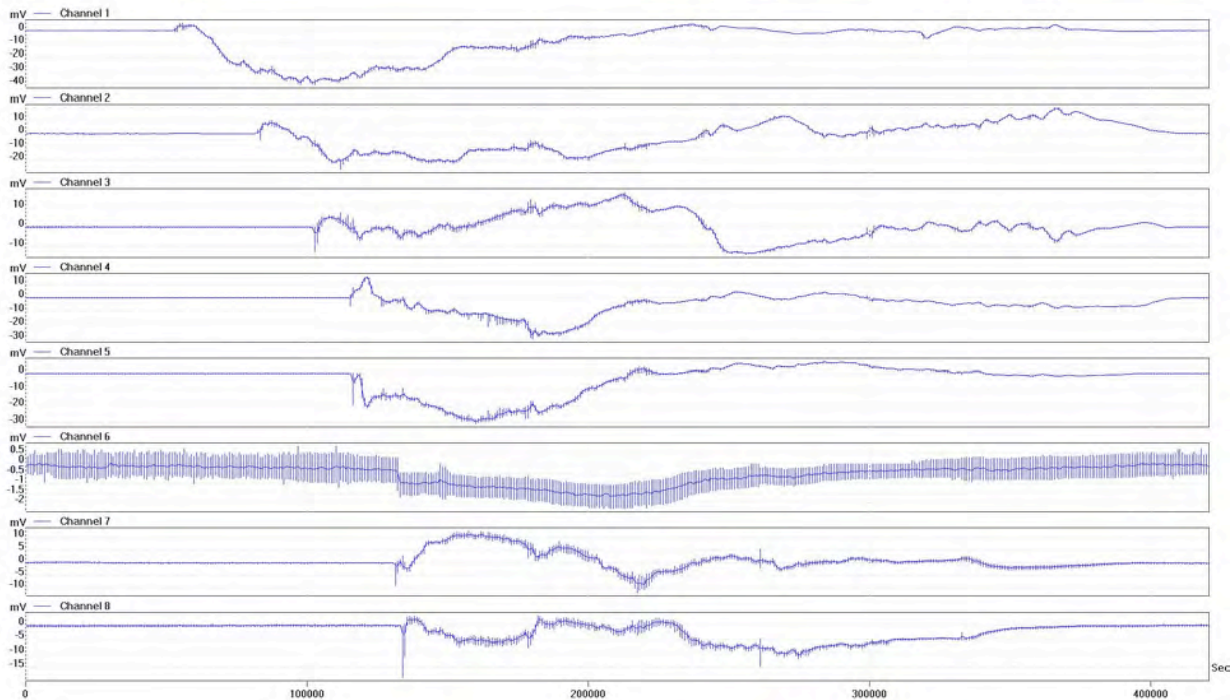

**Figure 4:** Plotting of the electric potentials for the 8 measurement electrodes. Note that electrode 6 (referred to as channel 6) registers only white noise because it has not been colonised by the plasmodium.

The behaviour of the plasmodium can be steered with light and substances that forges attraction and repulsion towards the electrodes. There are a number of substances that can act as attractants (e.g., carbohydrates and glucose) or repellent (e.g., potassium chloride and metal ions). However once these substances are added to the substrate they produce attracting or repelling gradients that are difficult to remove. Conversely, light does not present this problem, and therefore it is very suitable to control the plasmodium's behaviour dynamically.

*Physarum polycephalum* exhibits negative phototaxis. That is, the plasmodium moves away from light. In the presence of light at a certain spot (e.g., focused on one of the agar blobs) it either switches to another phase of its life cycle (by moving away from the blob) or undergoes fragmentation. Fragmentation normally occurs when the plasmodium cannot move away from the light (Guttes *et al*. 1961, Hilderbrandt 1986).

Blue or white light changes the plasmodium's oscillatory activity (Bailczyk 1979) and the closer the plasmodium is to the source of light, the stronger the influence of the light on the plamodium's oscillatory activity (Wohlfarth-Bottermann and Block 1981). Experiments conducted by Nakagaki and colleagues (1999) demonstrated that such oscillations could be synchronized with periodic illumination. These findings, and our own experiments in controlling plasmodium propagation with light, demonstrated that varying illumination gradients are good means to tune the plasmodium to produce specific oscillatory behaviours (Adamatzky 2009c).

### 4 The Sound Synthesis Engine

In order to render the plasmodium's voltages into sounds, we implemented an additive granular synthesiser (Miranda 2002). Granular synthesis works by generating a rapid succession of short sound bursts referred to as *sound granules* that together form larger sound events (Figure 5).

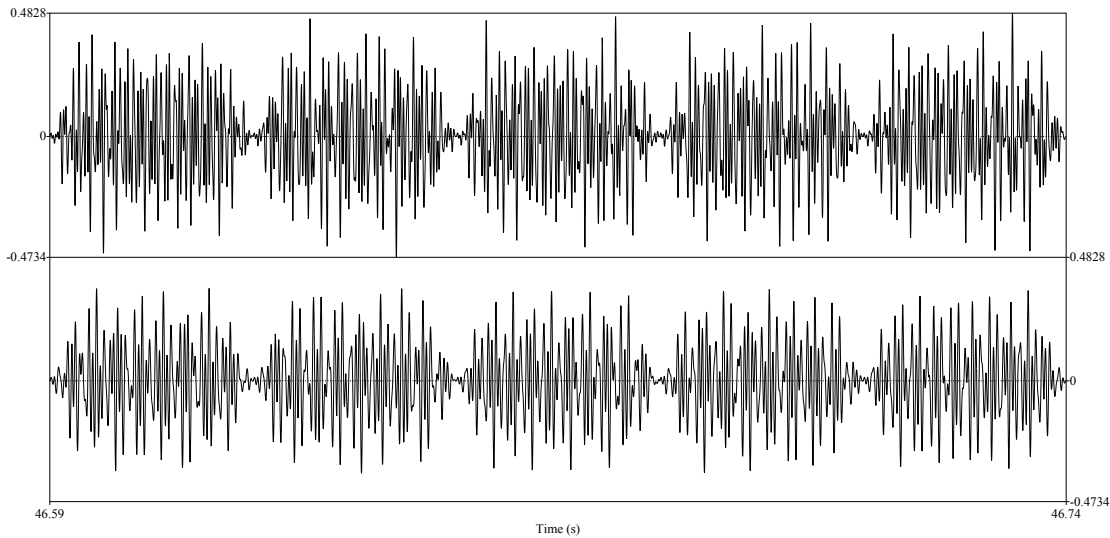

**Figure 5:** The sound synthesis engine is an additive granular synthesiser. In this example, 5 stereo sound granules each lasting for 30 ms, form a sound, which is 150 ms long.

Each sound granule is composed of $N$ spectral components, or partials, each of which is associated to a different measuring electrode, or channel. Thus, in the case of the example shown in Figure 5, each granule is composed of 8 sound partials. As the granules are in stereo, odd electrodes produce the left channel of sound and even electrodes the right one. Each sound partial is a sinewave produced by a digital oscillator, which needs two parameters to function: frequency and amplitude (Figure 6). (Phase information is sometimes needed, but we do not use phase information here.)

The voltages from the electrodes control either the frequencies of the oscillators or their frequencies and amplitudes together. In the first case, the voltages are normalized to a frequency range, which is set arbitrarily; e.g., between 20 Hz and 4 kHz. In this case, the amplitudes for each of the sine waves are fixed. In the second case, the voltages also control the amplitudes of the sine waves; here the voltages are also normalized to an amplitude range. In standard granular synthesis the duration of each granule is typically set in terms of tens of milliseconds. Such value can change dynamically as the sound is being synthesized. However, for the sake of simplicity of the examples mentioned in this paper, the granules are set to a fixed duration of 30 ms each.

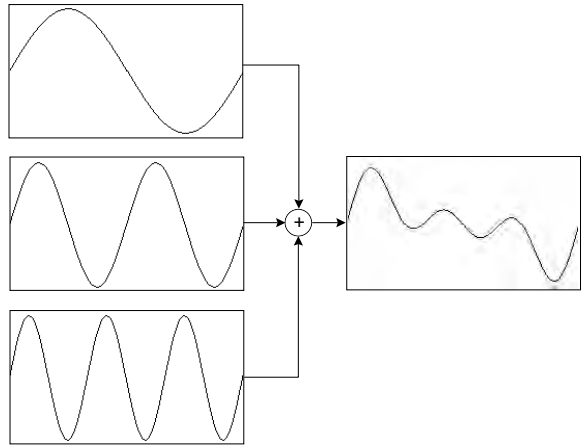

**Figure 6:** In additive synthesis each partial is a sinewave produced by an oscillator. The outputs of the oscillators are added together to form the resulting sound. In this example, the resulting sound results from the addition of 3 sinewaves at three different frequencies (speed of the oscillations), but identical amplitude (the height of the oscillation).

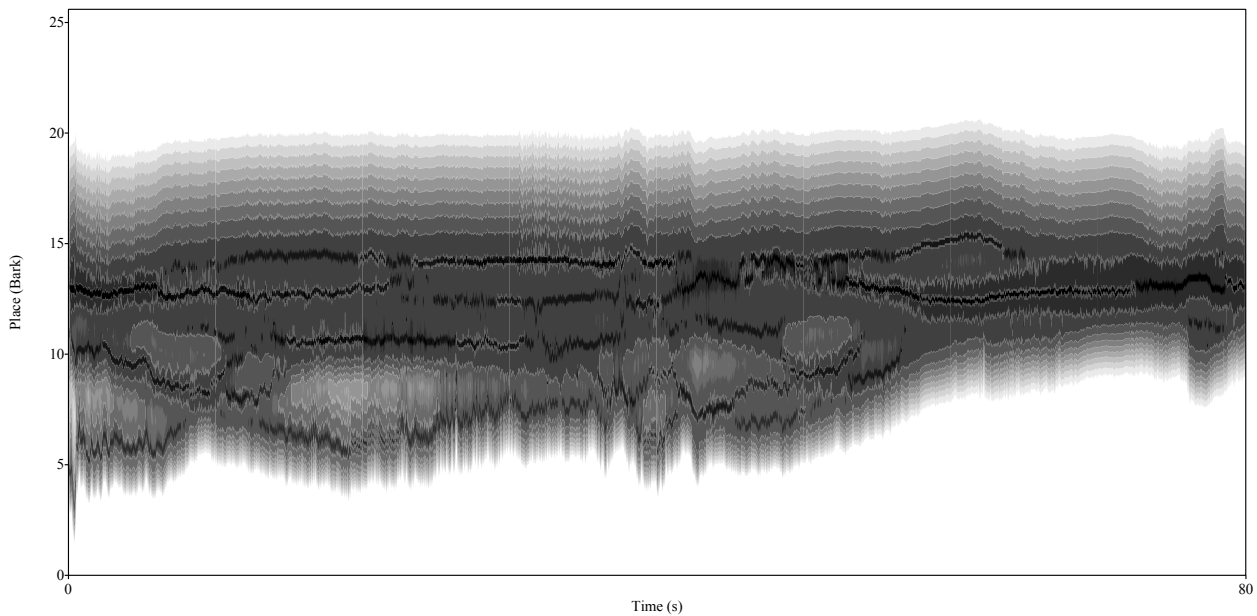

**Figure 7:** The cochleogram of a sound rendered from the data plotted in Figure 4.

Figure 7 shows the cochleogram of an 80 seconds-long sound rendered from the data plotted in Figure 4. In this case the voltages controlled only the frequencies of the oscillators.

Notice that despite the compression of the original raw data, as explained in section 2, there is a clear correspondence between the spectrum of the sound and the behaviour of the plasmodium. This is demonstrated by the darker lines of the cochleogram, which are morphologically related to the plotting of the voltages in Figure 4, even though both figures represent very different phenomena.

## 5 Simulation for Real-Time Synthesis

Obviously, the sound example in Figure 7 was generated off-line. The off-line version of the synthesiser can produce a variety of interesting sounds, which can be subsequently used by musicians in a number of different ways (e.g., they could be played back using a sampler or used in studio-based compositions). This is useful but the time it takes to run experiments with *Physarum polycephalum* can be tedious. Moreover, this poses a serious obstacle to implementing a synthesiser that could be played live as a musical instrument.

In order to address this problem, we utilise a computational approximation of *Physarum polycephalum* in the scenario we described in section 2 above. The model uses the multi-agent approach introduced in (Jones 2010) whereby simple, low-level interactions within a multi-agent collective generate emergent transport networks, which exhibit the network minimisation behaviours seen in *Physarum*. We use a modification of the model described in (Adamatzky and Jones, 2010) in order to cater for foraging, growth and adaptation behaviour of the plasmodium, and its reaction to attracting and repelling sources.

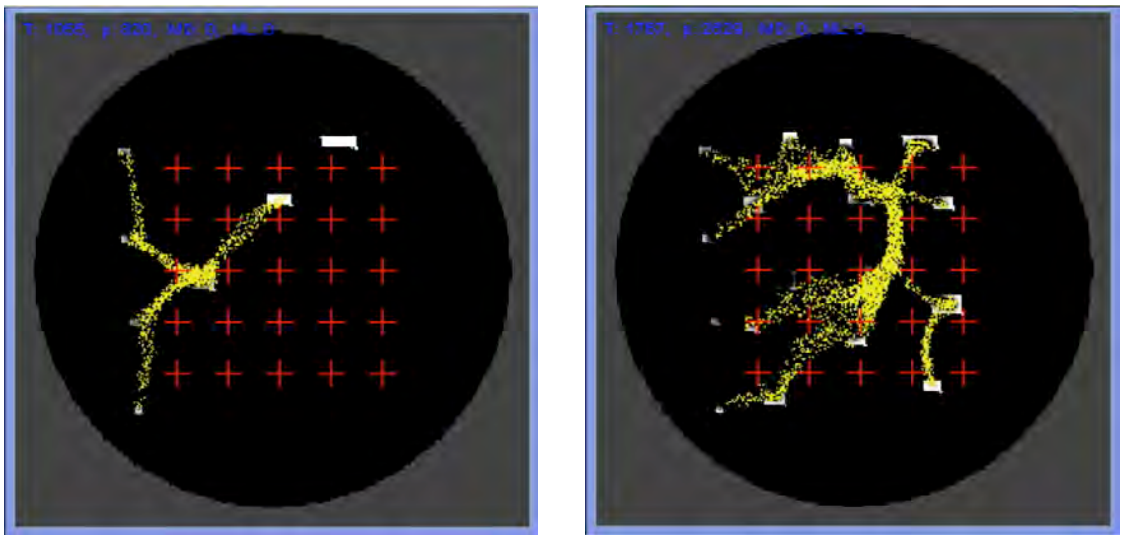

**Figure 8:** Two snapshots of a computer simulation of *Physarum polycephalum* foraging in a virtual Petri dish with an array of 25 virtual electrodes.

Figure 8 shows two snapshots of a simulation showing the plasmodium foraging in a virtual Petri dish with an array of 25 virtual electrodes; the number and the positions of electrodes we can place in the dish are arbitrary. Since the electrical activity of the plasmodium cannot be sampled within the model, we record the population size at every four steps of the scheduler. This was found to provide good spatial and temporal correlation with the electrical potential recordings (see Adamatzky and Jones, 2010 for further information). The electrodes are represented by crosses and the attractors are represented by squares. In this simulation one can add, consume and remove attractors inside the dish at will, therefore steering the behaviour of the plasmodium in real-time.

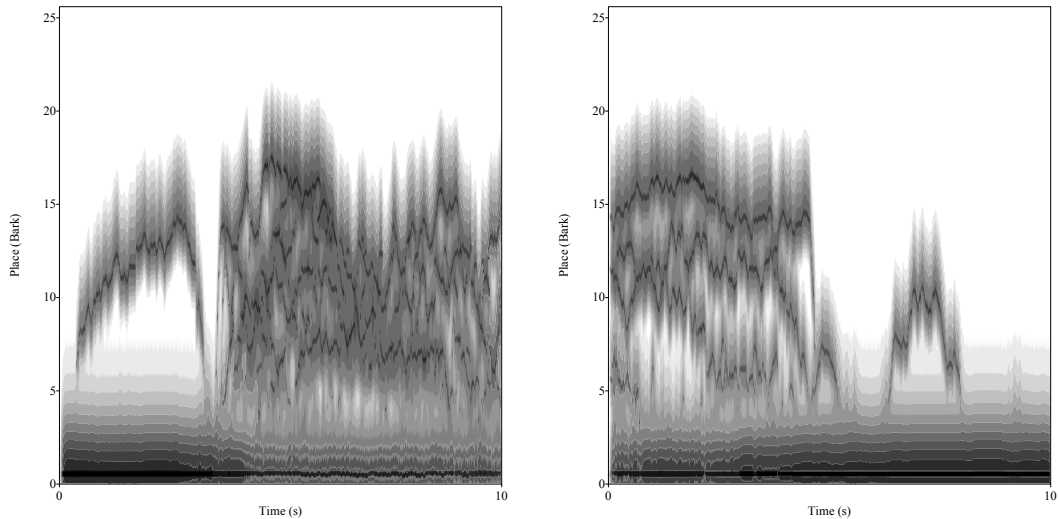

**Figure 9:** The cochleograms of two different portions of a sound produced with simulated data.

Figure 9 shows the cochleograms of two 10 second-long portions of a sound produced with simulated data, using the configuration shown in Figure 8. (But those cochleograms do not correspond to the two snapshots shown in Figure 8.) In this case, the voltages controlled both the frequency and amplitudes of the oscillators. On the left side of Figure 8 the spectrum became fatter as the plasmodium was steered to colonise more electrodes on the grid. And on the left side the spectrum became slimmer as the plasmodium was steered to move away from a number of electrodes on the grid.

**6 Conclusion**

In this paper we reported a method to render sounds from the electrical activity of *Physarum polycephalum*. The plasmodial slime mold *Physarum polycephalum* is a biological substrate used for research into unconventional computing.

At this stage of this research we are not concerned with studying the computational properties of the plasmodium. Rather, we are interested in studying its behaviour and how this behaviour can be rendered into sounds. *Physarum polycephalum* is interesting because it behaviour can be controlled to produce variations of its electrical activity (e.g., by placing attractors in the space) and consequently variations on the sounds.

The control of the behaviour of the plasmodium is still incipient and much research is currently under development by various laboratories worldwide. Also the speed of the plasmodium's behaviour makes it difficult to implement real-time synthesisers. Moreover, the data reduction needed to compress data worth of various days of plasmodium activity can hamper the relationship between the sound and the behaviour of the plasmodium. Of course, this relationship is not so important for music; a composer would not generally care if the sound corresponds exactly to what one would see in a graph plotting the raw data. Nevertheless, we still think that it is important to foster this relationship, because it allows for monitoring and prediction of the outcomes.

An interim solution to alleviate the limitations outlined above is to use a model that simulates the behaviour to the plasmodium. We have implemented such a model whereby we can simulate experiments much faster. Comparisons between real and simulated runs show that simulation data is sufficiently realistic.